\documentclass[twocolumn,eqsecnum,floatfix,aps,prl,nofootinbib]{revtex4}
\usepackage[dvips]{graphicx}
\usepackage{times}
\usepackage{cancel}

\def\fun#1#2{\lower3.6pt\vbox{\baselineskip0pt\lineskip.9pt
  \ialign{$\mathsurround=0pt#1\hfil##\hfil$\crcr#2\crcr\sim\crcr}}}

\usepackage[usenames]{xcolor}
\usepackage[normalem]{ulem}

\newcommand{\bec}[1]{\mbox{\boldmath $#1$}}

\begin{document}


\title{Possible lightest $\Xi$ Hypernucleus with Modern $\Xi N$ Interactions}

\author{E. Hiyama}
\affiliation{Department of Physics, Kyushu University, Fukuoka, Japan, 819-0395}
\affiliation{Strangeness Nuclear Physics Laboratory, RIKEN  Nishina Center,  Wako,  351-0198, Japan}

\author{K. Sasaki}
\affiliation{Center for Gravitational Physics, Yukawa Institute for Theoretical
Physics, Kyoto University, Kyoto 606-8502, Japan}

\author{T. Miyamoto}
\affiliation{Center for Gravitational Physics, Yukawa Institute for Theoretical
Physics, Kyoto University, Kyoto 606-8502, Japan}
\affiliation{Quantum Hadron Physics Laboratory, RIKEN Nishina Center, Wako,  351-0198, Japan}

\author{T. Doi}
\affiliation{Quantum Hadron Physics Laboratory, RIKEN Nishina Center, Wako,  351-0198, Japan}
\affiliation{Interdisciplinary Theoretical and Mathematical Sciences Program (iTHEMS), RIKEN, Wako,  351-0198, Japan}

\author{T. Hatsuda}
\affiliation{Interdisciplinary Theoretical and Mathematical Sciences Program (iTHEMS), RIKEN, Wako,  351-0198, Japan}
\affiliation{Quantum Hadron Physics Laboratory, RIKEN Nishina Center, Wako,  351-0198, Japan}

\author{Y. Yamamoto}
\affiliation{Physics Section, Tsuru University, Tsuru, Yamanashi 402-8555, Japan}
\affiliation{Strangeness Nuclear Physics Laboratory,RIKEN  Nishina Center,  Wako,  351-0198, Japan}

\author{Th. A. Rijken}
\affiliation{Institute for Theoretical Physics, University of Nijmegen, Njjmegen,
The Netherlands}

\date{\today}
%
\begin{abstract}

 Experimental evidence exists that
the $\Xi$-nucleus interaction is attractive.
We search for $NN\Xi$ and $NNN\Xi$ bound systems
on the basis of the  AV8 $NN$
potential combined with either a phenomenological
Nijmegen $\Xi N$ potential or a first principles HAL
QCD $\Xi N$ potential.
The binding energies of the
three-body and four-body systems (below the $d+\Xi$ and
$^3{\rm H}$/$^3{\rm He}+\Xi$ thresholds, respectively) are calculated by  a high precision variational approach,
 the  Gaussian Expansion Method. 
 Although  the two $\Xi N$ potentials have significantly different isospin ($T$) and spin ($S$) dependence,  
 the $NNN\Xi$ system with quantum numbers $(T=0, J^{\pi}=1^+$) appears to be bound 
 (one deep for Nijmegen
  and one shallow for HAL QCD) below the  $^3{\rm H}$/$^3{\rm He}+\Xi$ threshold.
  Experimental implications for such a state are discussed.

      \end{abstract}
 
\maketitle

One of the major goals of hypernuclear physics is
to understand the properties of  hyperon-nucleon ($YN$) and hyperon-hyperon ($YY$) interactions;
 they are related not only to  possible di-baryon states such as $H$ \cite{Jaffe} but also to the 
 role of hyperonic matter in neutron stars.
   Unlike the case of the  $NN$ interactions,
 hyperon interactions are not well determined   experimentally due to insufficient number of scattering data.
 Nevertheless,  high-resolution $\gamma$-ray experiments~\cite{Tamura2000,Ajimura2001,Akikawa2002,Ukai2004}
analyzed by the shell model~\cite{Millener1985} as well as the accurate few-body  method~\cite{Hiyama2009} have provided valuable
constraints  on the $YN$ interaction  in the ${\rm Strangeness}=-1$ sector such as  the $\Lambda N$ force.
 Also, the $\Lambda \Lambda$ interaction in the ${\rm Strangeness}=-2$ sector receives some constraints from
 the  binding energies of hypernuclei such as
 $^6_{\Lambda \Lambda}$He~\cite{Takahashi},
$^{10}_{\Lambda \Lambda}$Be~\cite{Danysz} and $^{13}_{\Lambda \Lambda}$B
~\cite{Aoki}.
 In addition, the femtoscopic analyses of the two-particle correlations in
  high-energy $pp$, $pA$ and $AA$ collisions at RHIC~\cite{STAR}
and LHC~\cite{Acharya:2018gyz,Acharya:2019yvb}  have started to give information on the  
 low-energy $\Lambda \Lambda$ scattering parameters. 

Recently, 
 the KEK-E373 experiment showed 
 a first evidence of a bound  $\Xi^-$ hypernucleus,
$^{15}_{\Xi}$C ($^{14}$N$+\Xi$), the "KISO" event~\cite{Kiso},
 which provides  useful information on the attractive $\Xi N$ interaction in the ${\rm Strangeness}=-2$ sector.  
It was suggested experimentally two possible $\Xi$ binding energies $B_\Xi \equiv E(^{15}_{\Xi}{\rm C})-E(^{14}{\rm N})$:
  $4.38 \pm 0.25$ MeV and $1.11 \pm 0.25 $ MeV.  
 The latest femtoscopic data from $pA$ collisions at LHC~\cite{Acharya:2019sms}
 also indicate that the spin-isospin  averaged $\Xi N$ interaction is attractive at low energies. 

Motivated by the above observations on  the $\Xi N$ interaction,
 we address the following questions in this Letter;
(i) what would be the lightest bound $\Xi$ hypernucleus? 
and  (ii) which  $\Xi N$ spin-isospin channel is responsible for such a bound  system?
In particular, we consider  three-body $NN\Xi$ and four-body $NNN\Xi$  systems
simultaneously using  a high-precision Gaussian Expansion Method (GEM)~\cite{Hiyama2003,Hiyama:2012sma},
which is one of the most powerful first principle methods
to solve three- and four-body problems.
We employ
  two modern $\Xi N$ interactions, 
  a phenomenological potential based on the meson exchanges, the Nijmegen $\Xi N$ potential (ESC08c) ~\cite{ESC08c},
  and a potential based on first principle lattice QCD simulations,  the HAL QCD $\Xi N$ potential (HAL QCD)~\cite{Sasaki:2019qnh}.
   As explained below, these two potentials have significantly different spin-isospin dependence.
  For the  $NN$ potential, we use the AV8 potential~\cite{Wiringa84}
 throughout this Letter.

 In the following, we employ the spectroscopic notation $^{2T+1,2S+1} {\rm S}_J$ to classify 
 the S-wave $\Xi N$ interaction
  where $T$, $S$ and $J$ stand for total isospin, total spin, and total angular momentum. 
  Thus we have four channels to be considered,  $^{11}{\rm S}_0$, $^{13}{\rm S}_1$, $^{31}{\rm S}_0$ and $^{33}{\rm S}_1$.
  As shown below, the largest attraction is in $^{33}{\rm S}_1$ and $^{11}{\rm S}_0$ for
     ESC08c and HAL QCD, respectively.

  Before entering the detailed  discussions on the three- and four-body systems,
   let us first summarize  key features of our $\Xi N$ potentials.
The ESC baryon-baryon potential is designed to describe $NN$, $YN$ and $YY$
 interactions in a unified way~\cite{ESC16}.
  In its recent version of ESC08c~\cite{ESC08c}, 
  a  $\pi \omega$-pair exchange potential $V_{\pi \omega}$ 
  is introduced  so as  to provide extra attraction
in the $T=1$ $\Xi N$ channel and  to be consistent with 
 the attractive nature  of $\Xi$-nucleus potential indicated by the $(K^-,K^+)$ 
experiments~\cite{E885} and the KISO event ~\cite{HYS2016}.
  Due to strong ($\Xi N$-$\Lambda \Sigma$-$\Sigma \Sigma$) central+tensor couplings
   in the $^{33}{\rm S}_1$ channel,  a $\Xi N$ (deuteron-like) bound state, $D^*$,
    is generated in ESC08c. (The $\Xi N$-$\Xi N$ sector composed of central and tensor terms is also attractive
but is not sufficient to form a two-body bound state.)
The $^{13}{\rm S}_1$ channel is weakly attractive, and
the $^{11}{\rm S}_0$ and  $^{31}{\rm S}_0$ channels are, on the other hand, repulsive in ESC08c.
 In this Letter,  we represent the ESC08c  by a $\Xi N$-$\Xi N$
single-channel potential with central and tensor components:
In  the $^{33}{\rm S}_1$ channel, the 
$\Xi N$-$\Lambda \Sigma$-$\Sigma \Sigma$ coupling effects are renormalized
into a $\Xi N$-$\Xi N$ central potential by adding a single-range Gaussian
form $V_2 \cdot \exp(-(r/\beta)^2)$ with  $V_2=-233$ MeV and 
$\beta$=1.0 fm.

The HAL QCD potential is obtained from first principles (2+1)-flavor 
lattice QCD simulations 
   in a large spacetime volume, $L^4 = (8.1 \ {\rm fm})^4$,  with nearly physical quark masses,
   $(m_{\pi}, m_K)$=(146, 525) MeV, at a lattice spacing, $a$=0.0846 fm.  
   Such simulations together with the HAL QCD method~\cite{Ishii:2006ec,HALQCD:2012aa} 
   enable one to extract the 
  $YN$ and $YY$ interactions with multiple strangeness, e.g. 
   $\Lambda\Lambda$, $ \Xi N$~\cite{Sasaki:2019qnh}, $ \Omega N$~\cite{Iritani:2018sra}  and
    $\Omega\Omega$~\cite{Gongyo:2017fjb}.

      We calculate the $\Xi N$ effective central interactions
      at the imaginary-time distances $t/a=11,12,13$,
      in which coupled-channel effect from higher channels as $\Lambda\Sigma$, $\Sigma\Sigma$
      are effectively included,
      whereas the effect from the lower channel ($\Lambda\Lambda$ in the $^{11}$S$_0$ channel)
      is explicitly handled by the coupled-channel formalism~\cite{Aoki:2011gt,Aoki:2012bb}.

  To make the few-body calculation feasible, we fit  the lattice QCD result of the 
 potentials
    with multiple  Gaussian forms at short distances and the Yukawa  form  
    with  form factors at medium to long distances~\cite{Sasaki:2019qnh}.  As for the pion and Kaon masses which 
     dictate the long range part of the potential, we use 
     $(m_{\pi}, m_K) = (146, 525)$ MeV to fit the 
      lattice data, and take 
      $(m_{\pi}, m_K) = (138, 496)$ MeV 
      for calculating the $\Xi$-nucleus systems.
   In the $^{11}{\rm S}_0$ channel,
    the analysis of the $\Lambda\Lambda$ and $N \Xi$ scattering phase shifts
  shows  that  a 
  $\Xi N$ interaction is moderately attractive.
 Also, deeply bound $H$-dibaryon is not found below the $\Lambda\Lambda$ threshold.
  Moreover, the channel-coupling
      between  $\Lambda\Lambda$ and $\Xi N$ is found to be  weak~\cite{Sasaki:2019qnh}.
         On the basis of these evidences, 
         we  introduce an effective single-channel $\Xi N$ potential in which 
   the coupling to $\Lambda\Lambda$  in $^{11}${\rm S}$_0$  is renormalized into  
   a single range Gaussian form   
$U_2 \cdot {\rm exp}(-(r/\gamma)^2)$ with $\gamma$=1.0 fm with $U_2 (<0)$ chosen to reproduce the 
 $\Xi N$ phase shifts obtained with channel coupling.
  On the other hand, 
  the $\Xi N$ interactions in other channels 
  are found to be much weaker: The $^{13}$S$_1$ and $^{33}$S$_1$ channels are weakly attractive and
  the $^{31}$S$_0$ channel is weakly repulsive.

\begin{figure*}[htb]
\begin{center}
\begin{minipage}{0.45\hsize}
\includegraphics[scale=0.4]{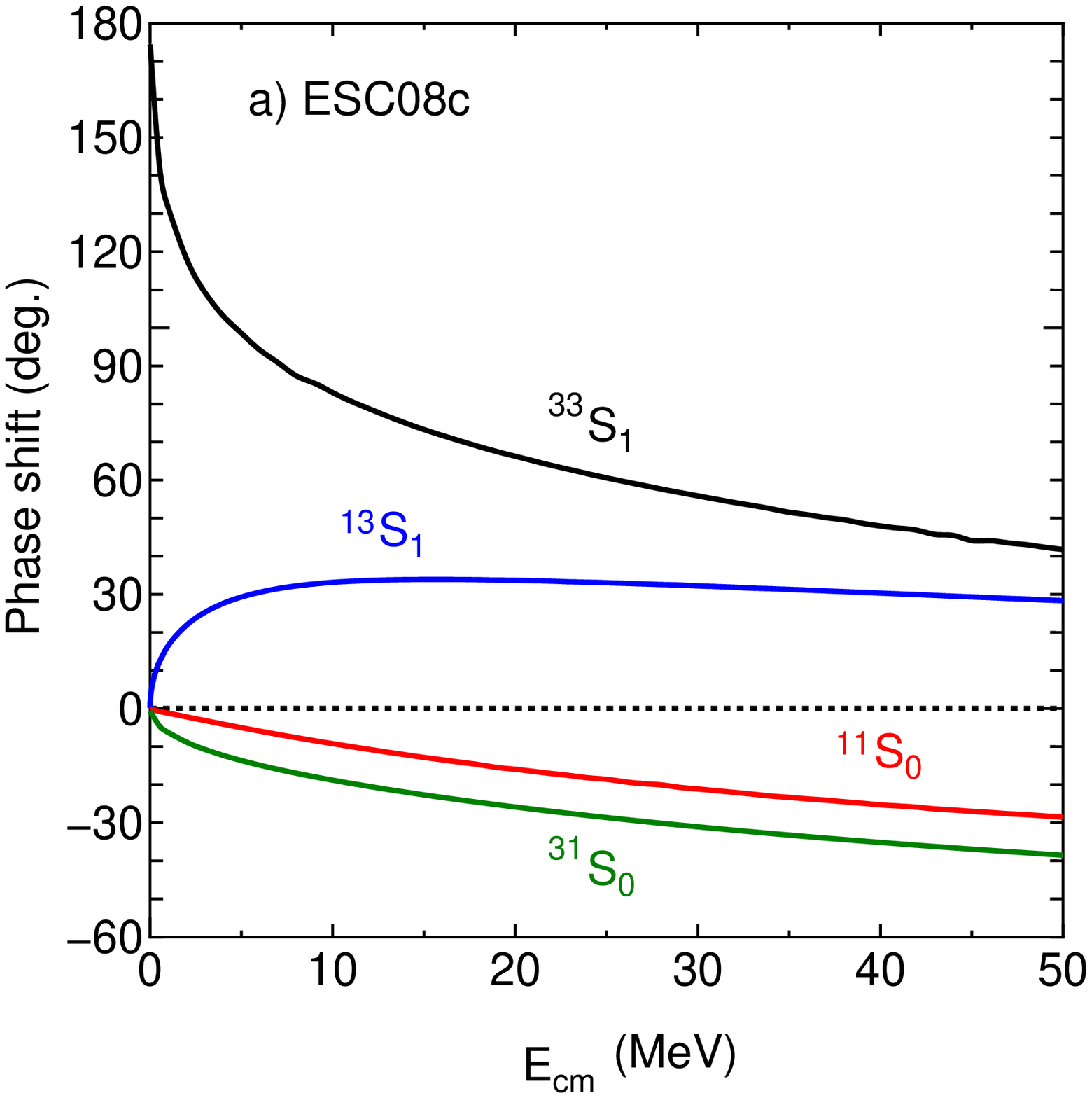}
\end{minipage}
\begin{minipage}{0.45\hsize}
\includegraphics[scale=0.4]{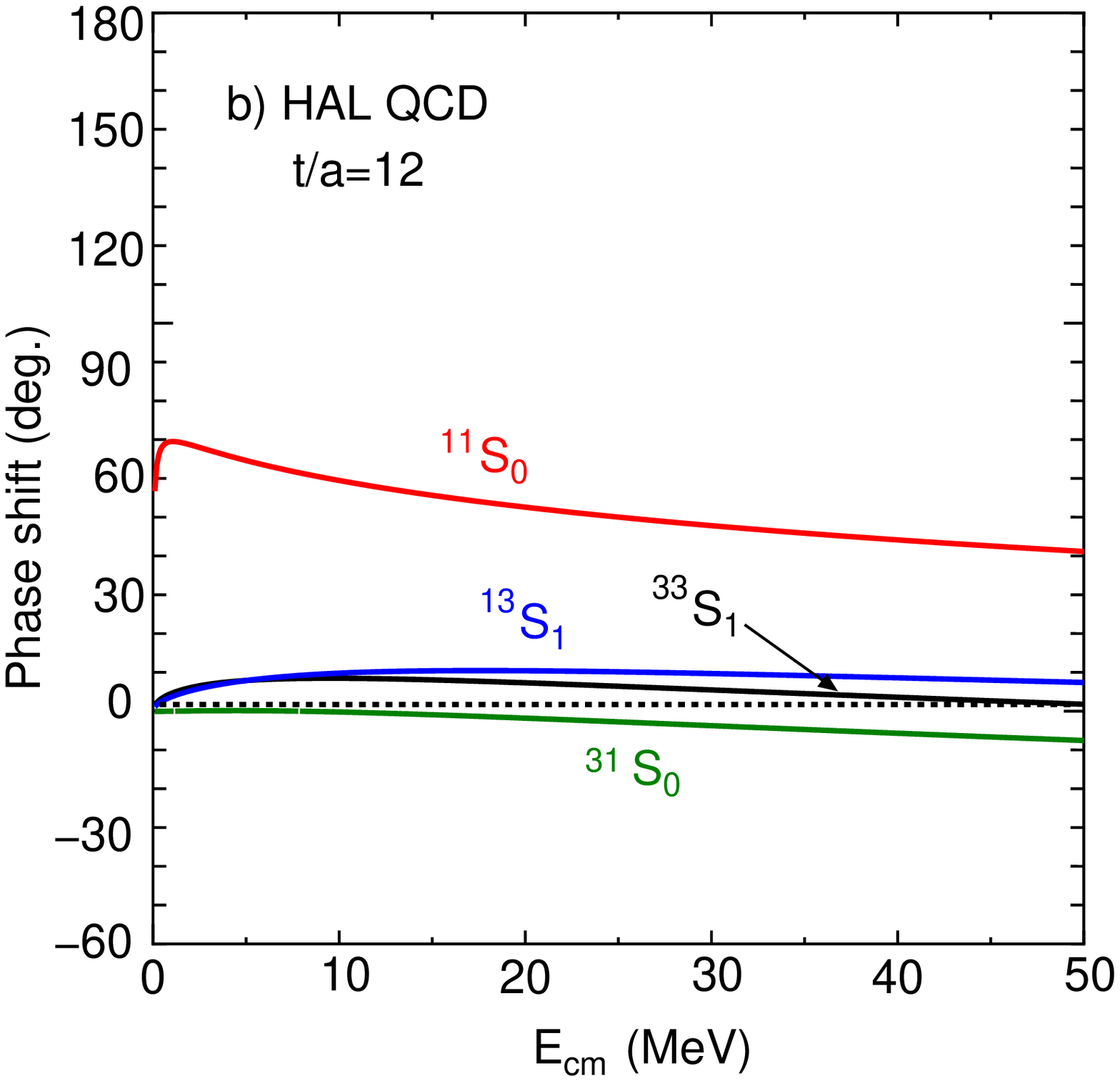}
\end{minipage}
\end{center}
\caption{$\Xi N$ phase shifts in the $^{33}{\rm S}_1$, $^{13}{\rm S}_1$, $^{11}{\rm S}_0$ and $^{31}{\rm S}_0$ channels using
(a) the ESC08c potential and (b) the HAL QCD potential ($t/a=12$).
}
\label{fig:phase}
\end{figure*}

In Fig.~\ref{fig:phase}, we show  the $\Xi N$ phase shifts calculated with  (a) the ESC08c potential and (b) the 
 HAL QCD potential at $t/a=12$ for comparison.   The  statistical and systematic errors 
  are not shown in Fig.~\ref{fig:phase}(b),  but are taken into account in the few body calculations below.
   From the figure, one immediately finds a qualitative difference  between  (a) and (b):
    The $^{33}{\rm S}_1$ channel is   attractive  in  ESC08c even to form a bound state with the binding energy of 1.59 MeV,
     while it  has only weak attraction  in HAL QCD.
  On the other hand,  the $^{11}${\rm S}$_0$ channel is repulsive  in  ESC08c, while it  is moderately attractive in HAL QCD.
  It is therefore  interesting  to see how such differences
 are reflected in the energy levels of the  few-body $\Xi$ hypernuclei.

In this Letter, we consider the $NN\Xi$ and $NNN\Xi$ systems simultaneously
 by using GEM ~\cite{Hiyama2003,Hiyama:2012sma}.~\footnote{We note that  
 the $NN\Xi$  system was recently studied by 
the Faddeev method~\cite{Garcilazo}  with an effective  $\Xi N$ potential 
 inspired by ESC08c.
 Two bound states are found 
$B_{\Xi}= 13.5$ MeV with $(T, J^\pi)=(1/2, 3/2^+)$  and
 $B_{\Xi}=0.012$ MeV with  $(T, J^\pi)=(1/2, 1/2^+)$
with respect to the $d + \Xi $ threshold.
In addition,
 one bound state is found to be
  $1.33$ MeV with   $(T, J^\pi)=(3/2, 1/2^+)$
  with respect to the $D^* + N$ threshold.}

GEM is a variational method
with Gaussian bases, 
which achieves similar accuracy for bound state problems
to
other methods such as
Faddeev method and
Green Function Monte Carlo method~\cite{Kamada:2001tv}.
GEM has been applied successfully up to five-body problems.

 For ordinary nuclei without strangeness, we will not consider the isospin breaking
 from strong interaction nor the Coulomb interaction, so that $T$ is a good quantum number.
  For  the $N\Xi$ interaction, however, we take into account both strong interaction and
   the Coulomb interaction, since the latter effect may not be negligible for  weakly bound $\Xi$-nuclei.
 Accordingly possible isospin breaking such as  the mixing between $T=0$ and $T=1$ for $NNN\Xi$ may occur.
  
\begin{figure}[htb]
\centerline{
\includegraphics[scale=0.4]{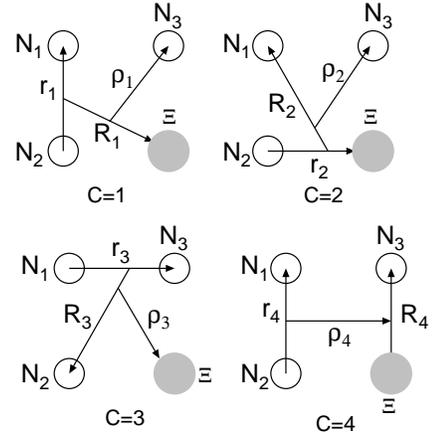}}
\caption{Jacobi coordinates for the rearrangement channels of the $NNN\Xi$ system.}
\label{fig:jacobi}
\end{figure}

  In GEM, three and four  Jacobi coordinates are introduced to describe  $NN\Xi$ and $NNN\Xi$, respectively.
  Shown in Fig.~\ref{fig:jacobi} are the four rearrangement  channels in 
   $NNN\Xi$. The  four-body wavefunction is given as a sum of 
 $c=1 \sim 4$ in  Fig.~\ref{fig:jacobi} with the LS coupling scheme:
\begin{eqnarray}
\Psi_{JM}
 &=& \sum^4_{c=1} \sum_{\alpha I}
\sum_{ss'Stt'T} C^{(c)}_{\alpha Iss'Stt'T}  \nonumber \\
& & {\cal A} \left[ \left[ \phi_{\alpha I }^{(c)} (\bec{r}_c, \bec{R}_c, \bec{\rho}_c)
[[\chi_s(12)\chi_{\frac{1}{2}}(3)]_{s'}\chi_{\frac{1}{2}}(\Xi)]_S \right]_{JM} \right. \nonumber \\
& &\ \ \ \ \ \cdot  \left.[ [\eta_t(12) \eta_{\frac{1}{2}}(3)]_{t'}\eta_{\frac{1}{2}}(\Xi) ]_{T,T_z} \right] . 
\end{eqnarray}
Here $\cal{A}$ denotes anti-symmetrization operator for the nucleons.
  Spin and isospin functions are denoted by $\chi$'s and 
$\eta$'s, respectively. Total isospin $T$ can in principle take the values
 $0,1,2$.  However,  $T=2$ corresponds to the 3N state of $t'=3/2$ in the continuum, so that
  its contribution is negligible. The spatial wavefunctions have the form,
$\phi_{\alpha I M'}(\bec{r, R, \rho})=
[[\phi_{n\ell}(\bec{r})\psi_{NL}(\bec{R})]_K \xi_{\nu \lambda}(\bec{\rho})]_{I M'}$
with a set of quantum numbers,
$\alpha =(n, \ell; N, L; K; \nu, \lambda)$, and
the radial components of
  $\phi_{n\ell m}(\bec{r})$ are taken as
$r^\ell e^{-{(r/r_n)}^2}$,
where the range parameters $r_n$  are chosen
to satisfy a geometrical progression.
Similar choice for  $\psi_{NL}(\bec{R})$ and  $\xi_{\nu \lambda}(\bec{\rho})$ are taken.
These four-body basis functions  are known to be sufficient for
describing both the short-range correlations and the long-range
tail behavior of the few-body systems.
 The 3N binding energy with the present AV8 $NN$ potential
  becomes $7.78$MeV which is less than the observed binding energy   $8.48$ MeV of $^3$H.
This discrepancy is attributed to the three-body force, so that  
 a phenomenological attractive three-body potential defined by
$W_3 \cdot \exp( {- \sum_{i>j} (r_{ij}/\delta)^2}$)
 is introduced, where $r_{ij}$
 are the relative distances between the three nucleons $N_{i}$,
  with $W_3=-45.4$ MeV and  $\delta=1.5$ fm.

\begin{table} [tbh] \begin{center} 
\caption{The calculated binding energies (in units of MeV)
 of  $NN\Xi$ and $NNN\Xi$ with
ESC08c potential and with HAL QCD potential with respect to the $d+\Xi$ and 
$^3{\rm H}/^3{\rm He}+\Xi$ threshold, respectively.
}
\label{table:result-energy}
\begin{tabular}{|c|cc|cccc|}
\hline
                   &  $NN\Xi$     &                       &                                      &  $NNN\Xi$          
                      &                 &  \\
\hline \hline
$(T,J^{\pi})$    &  $(\frac{1}{2}, \frac{1}{2}^+)$ &  $(\frac{1}{2}, \frac{3}{2}^+)$   &  $(0, 0^+$)    & $(0, 1^+$)       & $(1, 0^+$)  &  $(1, 1^+)$   \\
\hline
ESC08c               & $-$                   &  $7.20$     & $-$                &  $10.20$               &    $3.55$                & $10.11$    \\
HAL QCD            & $-$                   &  $-$           & $-$                &  $0.36 (16)(26)$      &    $-$                     &  $-$           \\
\hline
\end{tabular}
\end{center}
\end{table}

  In TABLE I, we summarize the binding energies of $NN\Xi$ and $NNN\Xi$ systems,
  where we omit atomic states which are (almost) purely bound by the Coulomb interaction.
We note that the isospin mixing by the Coulomb interaction 
 is found to be small, so that the states can be labeled by $T$ in good approximation.

Let us now discuss the results
 with the  ESC08c $\Xi N$ potential.
 The  binding energy of the  $NN \Xi$ system with
$(T,J^{\pi})=(1/2, 3/2^+)$  with respect to the $d+\Xi$ threshold
 is 7.20 MeV, while the $NN \Xi$
with $(T,J^{\pi})=(1/2, 1/2^+)$  is unbound.
Such channel dependence can be easily understood  in the following manner:
 For $NN\Xi (1/2, 3/2^+)$, nucleon and $\Xi$ spins  are all aligned.
  Since the nuclear force is most attractive in the spin-1 pair, and 
  the $\Xi N$ force in ESC08c is also attractive for spin-1 pairs as shown in Fig.1 (a),
  this channel is most attractive to bring the bound state.
 On the other hand,  in  $NN\Xi (1/2, 1/2^+)$,  one of the nucleon spins  or $\Xi$ spin 
 is anti-parallel to the  others, so that  one or two spin-0 $\Xi N$ pair
 appear in the wave function. Since such pair is repulsive in ESC08c
 as shown in Fig.1 (a), this channel becomes  unbound.
Note here that our results of $NN\Xi$ are qualitatively similar to
 but numerically different  from  those in~\cite{Garcilazo}
   due to different  $NN$ potential  
 and different  treatment of  ESC08c.
  In $T=3/2$ $NN\Xi$ channel , we do not find a bound state with respect to the $D^* + N$ threshold,
  while one bound state is found with $(3/2, 1/2^+)$ in~\cite{Garcilazo}.

For  $NNN\Xi$ system  in ESC08c,
 the state in $(T,J^\pi)=(0,0^+)$  is  unbound with respect to $^3{\rm H}/^3{\rm He}+\Xi$ 
threshold, while the states in $(T,J^\pi)= (0,1^+), (1,0^+)$ and $(1,1^+)$  
are bound by  $10.20, 3.55$ and  $10.11$ MeV, respectively, as shown in TABLE I. 
The effect of the  $\Xi N$ Coulomb
interaction to these binding energies are only 10-20\% of those numbers.
  The physical reason behind such channel dependence is more involved than the 
 case of $NN\Xi$ due to various combinations of the pairs.
  Nevertheless, we find that
  the dominant $\Xi N$ pair in the $(T,J^\pi)=(0,0^+)$ system
  is the repulsive $^{11}{\rm S}_0$ channel in ESC08c, which leads to  the unbinding of this system.
On the other hand, the dominant $\Xi N$ pairs
in $(T,J^\pi)=(1,1^+)$ and $(0,1^+)$ systems
are $^{33}{\rm S}_1$ and $^{13}{\rm S}_1$ channels so that
the binding energies of these $NNN\Xi$ systems are large.

Let us now turn to the $NN\Xi$ and $NNN\Xi$ systems
 with  the HAL QCD $\Xi N$  potential.
 We found that none of the  potentials ($t/a=11,12$ and $13$)
 support bound states for $N\Xi$ and $NN\Xi$ systems.
  Only for the four-body $NNN\Xi$ system with $(T,J^{\pi})=(0,1^+)$, we have 
a  possibility of a shallow bound state with 
 the binding energies of $0.63 \ (t/a=11), 0.36 \ (t/a=12),  0.18\ (t/a=13)$ MeV 
 with respect to the  $^3{\rm H}/^3{\rm He}+\Xi$ threshold.
  In TABLE I, we quote the number {0.36 (16)(26) MeV where the first parenthesis
  shows the  error  originating from the statistical error of the $\Xi N$ potential at $t/a=12$
   and the second  parenthesis shows the systematic error. The former is
   estimated by the jackknife sampling of the lattice QCD configurations and the
    latter is estimated from the data at $t/a=11$ and $13$. 

 The reason why the bound state is so shallow is that, unlike the case of ESC08c,
  the  HAL QCD potential  is moderately attractive in  $^{11}{\rm S}_0$, while 
  it  is  either weakly attractive or repulsive in other channels as shown in Fig.~\ref{fig:phase} (b).
  If we switch off the Coulomb interaction, the bound state at $t/a=12$ (and 13) disappears.
  Therefore, this is a Coulomb-assisted bound state.  However,
   the contribution from the strong $\Xi N$  interaction  is still substantially larger
  than that of Coulomb $\Xi N$ interaction as seen from their
   expectation values, $\langle  V_{\Xi N}^{\rm strong} \rangle =-2.06$ MeV vs.
   $\langle V_{\Xi N}^{\rm Coulomb}\rangle =-0.38$MeV for $t/a=12$.
   Also, the mixing of the $(T,J^{\pi})=(1,1^+)$ state to the $(T,J^{\pi})=(0,1^+)$ state
    due to Coulomb effect is less than 1\% for $t/a$=12.

\begin{figure*}[htb]
\begin{minipage}{0.45\hsize}
\includegraphics[scale=0.4]{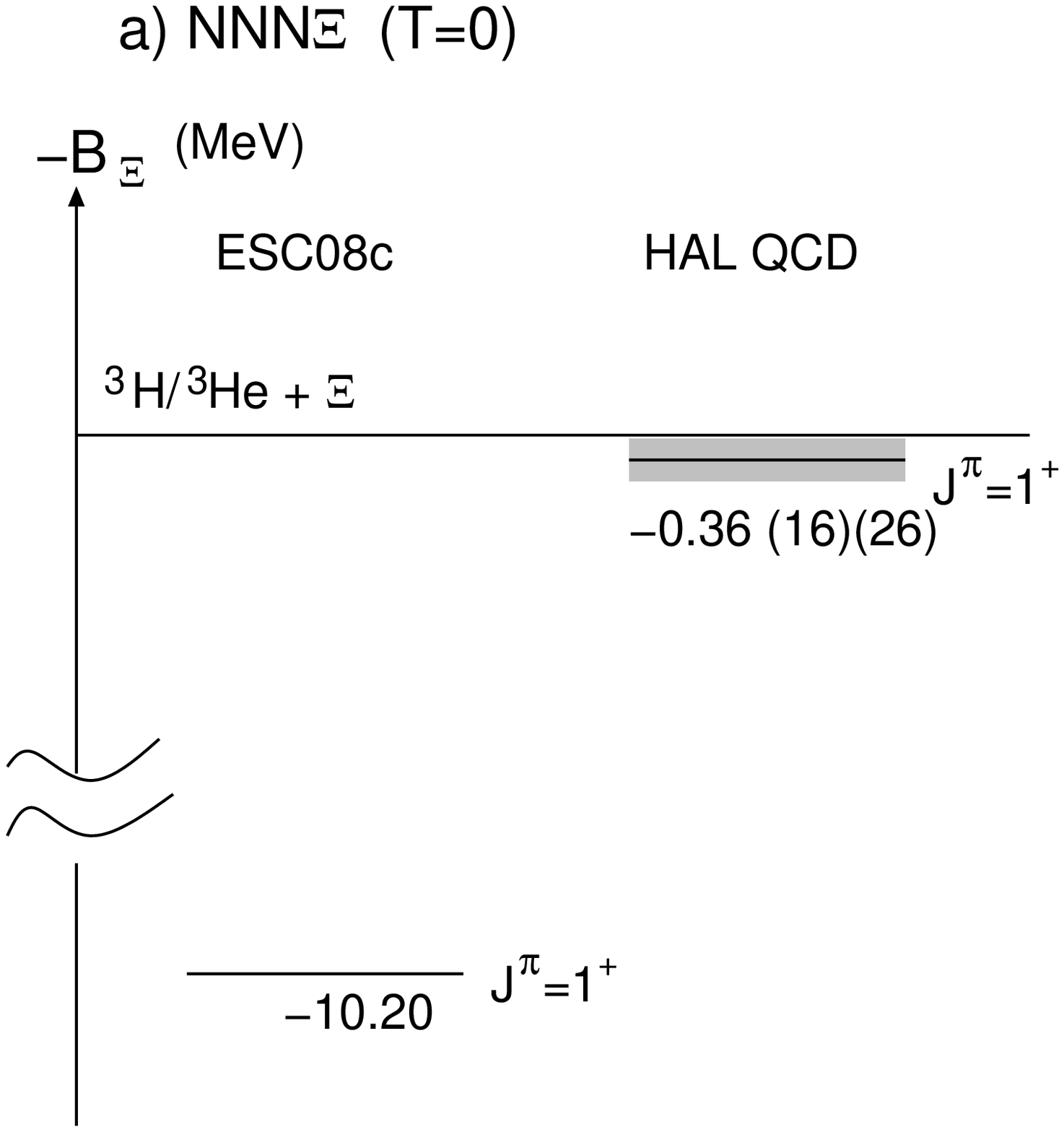}
\end{minipage}
\begin{minipage}{0.45\hsize}
\includegraphics[scale=0.4]{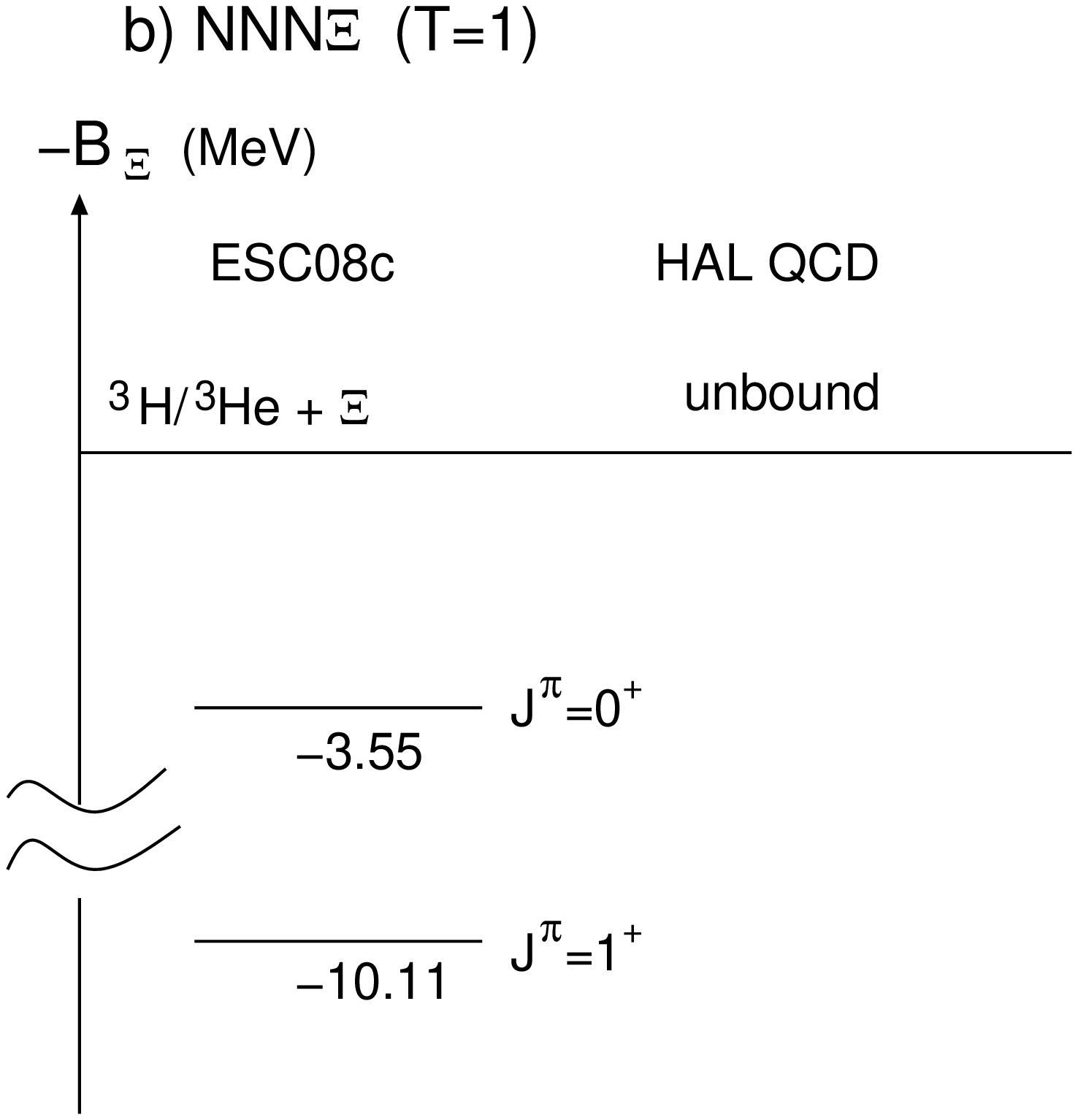}
\vspace*{1.0em}
\end{minipage}
\caption{Binding energies of $NNN\Xi$ system using ESC08c and HAL QCD
potentials for (a) $(T,J^{\pi})=(0,  1^+)$ and (b) $(T,J^{\pi})=(1,  0^+), (1,  1^+)$ states.
The gray band for HAL QCD is obtained by the quadrature of the statistical and systematic errors.
}
\label{fig:spectra}
\end{figure*}

Shown in Fig.3 is  a comparison of the $NNN\Xi$ binding energies calculated with 
ESC08c and HAL QCD.  In both cases, $NNN \Xi$ in $(T,J^{\pi})=(0,1^{+})$ (Fig.~\ref{fig:spectra} (a)) is 
  a possible candidate of the  lightest $\Xi$ hypernucleus.
 The binding energy  and the binding mechanism are, however, 
 totally different between the two cases;
 the  strong attraction in $^{33}{\rm S}_1$ drives $\sim 10$ MeV binding  for the ESC08c potential,
  while  the moderate attraction in $^{11}{\rm  S}_0$  leads to a binding less than 1 MeV for the 
   HAL QCD potential. 

Here, we note that all the $NNN\Xi$ states in Fig.~\ref{fig:spectra} are
  the resonant states
 above the $N+N+\Lambda+\Lambda$ threshold.
We estimate perturbatively the decay width $\Gamma$  of $NNN\Xi$  by using 
the $\Xi N$-$\Lambda \Lambda$ coupling potential and found that 
 $\Gamma =0.89, 0.43, 0.03$ MeV for $(0,1^+), (1,0^+), (1,1^+)$, respectively, with ESC08c.
With HAL QCD, $\Gamma=0.06, 0.05, 0.03$ MeV in $t/a=11,12,13$, respectively, for $(0,1^+)$.
 In both cases, the decay widths are sufficiently small for those states  to be observed.

To produce $NNN\Xi $ states  experimentally, 
  heavy ion reactions at GSI and CERN LHC would be useful. 
    If there exists a bound $NNN\Xi (0,1^+)$,  
 it decays into $d+\Lambda +\Lambda$ or a possible 
  double $\Lambda$ hypernucleus $^4_{\Lambda \Lambda}$H
 through the $\Xi N$-$\Lambda \Lambda$ coupling.
 On the other hand,  to produce $NNN\Xi (1,0^+)$ and 
 $NNN\Xi (1,1^+)$    states as predicted by ESC08c,
the $(K^-,K^+)$ reaction with a $^4$He target will be useful.

Finally, we remark that 
$^4_{\Lambda \Lambda}$H with $\Lambda \Lambda$-$\Xi N$
and $\Lambda N$-$\Sigma N$ couplings  has been studied before
 with phenomenological $YN$ and $YY$ interactions~\cite{Nemura05}.\footnote{
Here we note that they used the  observed data for the two-$\Lambda$ separation energy of
$^6_{\Lambda \Lambda}$He, $B_{\Lambda \Lambda}
=7.25\pm 0.19^{+0.18}_{-0.11}$ MeV. Afterwards, the revised data
 $B_{\Lambda \Lambda}=6.91 \pm 0.16$ MeV was reported, which implies that 
 the $\Lambda \Lambda$ attraction is slightly  weaker.}
 They reported  possible existence of a weakly bound state below $d+\Lambda
+\Lambda$ threshold, which has not yet been confirmed experimentally \cite{Ahn01}. 
 Also, Cottenssi {\it et al.} \cite{Cottenssi} have recently
emphasized that the particle stability of 
$A=5$ double $\Lambda$ hypernuclei ($^5_{\Lambda \Lambda}$He
and $^5_{\Lambda \Lambda}$H) is robust.
It is, therefore,  tempting to revisit the  
  $^4_{\Lambda \Lambda}$H system together with the $NNN \Xi (0,1^{+})$
  with modern coupled channel  baryon-baryon interactions
 to answer the  following question,  ``What would be the lightest 
Strangeness$=-2$ nucleus ?''.
The  analyses and the results of the present work provide a first step towards the 
 goal.

\acknowledgements
The authors would like to thank Prof. B.F. Gibson for useful discussions.
This work is supported in part
by JSPS Grant-in-Aid for Scientific Research
(No. JP18H05407, JP16H03995, JP18H05236, JP19K03879),
by a priority issue (Elucidation of the fundamental laws and evolution of the 
universe) to be tackled by using Post ``K" Computer,
and by Joint Institute for Computational Fundamental Science (JICFuS).
The authors thank HAL QCD Collaboration for providing lattice QCD results of $\Xi N$ interactions
and for valuable discussions.



\begin{thebibliography}{99}

\bibitem{Jaffe} R. L. Jaffe, Phys. Rev. Lett. {\bf 38}, 195 (1977).

\bibitem{Tamura2000} H. Tamura {\it et al.}, 
Phys. Rev. Lett. {\bf 84}, 5963 (2000).

\bibitem{Ajimura2001} S. Ajimura {\it et al.},
Phys. Rev. Lett. {\bf 86}, 4255 (2001).

\bibitem{Akikawa2002} H. Akikawa {\it et al.},
Phys. Rev. Lett. {\bf 88}, 082501 (2002), H. Tamura {\it et al.}, Nucl.
Phys. {\bf A 754} 58c (2005).

\bibitem{Ukai2004} M. Ukai {\it et al.} 
Phys. Rev. Lett. {\bf 93}, 232501 (2004), ibid.
Phys. Rev. C{\bf 73} 012501(R) (2006). 

\bibitem{Millener1985} 
A. Gal, E. V. Hungerford, and  D. J. Millener,  
Rev. Mod. Phys. {\bf 88}, 035004 (2016).

\bibitem{Hiyama2009} E. Hiyama and T. Yamada, 
Prog. Par. Nucl. Phys. {\bf 63}, 339 (2009).

\bibitem{Takahashi}
H. Takahashi {\it et al.}, 
Phys. Rev. Lett. {\bf 87}, 212502 (2001).

\bibitem{Danysz} 
M. Danysz {\it et al.},  
Phys. Rev. Lett. {\bf 11}, 29 (1963), ibid. Nucl. Phys. {\bf 49}, 121 (1963).

\bibitem{Aoki} 
S. Aoki {\it et al.},
Prog. Theor. Phys. {\bf 85}, 1287 (1991).


\bibitem{STAR} Adamczyk et al. [STAR collaboration],
Phys. Rev. Lett. {\bf 114} 022301 (2015).

\bibitem{Acharya:2018gyz} 
  S.~Acharya {\it et al.} [ALICE Collaboration],
  Phys.\ Rev.\ C {\bf 99},  024001 (2019)  [arXiv:1805.12455 [nucl-ex]].

\bibitem{Acharya:2019yvb} 
  S.~Acharya {\it et al.} [ALICE Collaboration],
  arXiv:1905.07209 [nucl-ex].
  
 \bibitem{Kiso} 
 K. Nakazawa {\it et al.}, Prog. Theor. Exp. Phys. {\bf 33}, D02 (2015).
 

\bibitem{Acharya:2019sms} 
S.~Acharya {\it et al.} [ALICE Collaboration],
  Phys.\ Rev.\ Lett.\  {\bf 123}, 112002 (2019)
  [arXiv:1904.12198 [nucl-ex]].


\bibitem{Hiyama2003}
E. Hiyama, Y. Kino, and M. Kamimura,
Prog. Theor. Nucl. Phys. {\bf 51}, 223 (2003).

\bibitem{Hiyama:2012sma} 
  E.~Hiyama,
  PTEP {\bf 2012}, 01A204 (2012).

\bibitem{ESC08c} 
M.M. Nagels, Th.A. Rijken, and Y. Yamamoto, 
arXiv:1504.02634 (2015).

\bibitem{Sasaki:2019qnh}

  K.~Sasaki {\it et al.} [HAL QCD Collaboration],
  arXiv:1912.08630 [hep-lat],
%
  and a paper in preparation.



\bibitem{Wiringa84} 
  R. B. Wiringa, R. A. Smith, and T. L. Ainsworth, 
  Phys. Rev. C {\bf 29}, 1207 (1984). 
 

\bibitem{ESC16} 
M. M. Nagels, Th. A. Rijken, and Y. Yamamoto,
Phys. Rev. C {\bf 99}, 044002 (2019); {\it ibid.} {\bf  99}, 044003 (2019).

\bibitem{E885}
P.~Khaustov {\it et al.}, Phys. Rev. {\bf C61}, 054603 (2000).

\bibitem{HYS2016} 
E. Hiyama, Y. Yamamoto, and H. Sagawa,
Physica Scripta, Volume 91, Number 9 (2016)

\bibitem{Ishii:2006ec} 
  N.~Ishii, S.~Aoki and T.~Hatsuda,
  Phys.\ Rev.\ Lett.\  {\bf 99}, 022001 (2007)
    [nucl-th/0611096].
    
\bibitem{HALQCD:2012aa} 
  N.~Ishii {\it et al.} [HAL QCD Collaboration],
  Phys.\ Lett.\ B {\bf 712}, 437 (2012)
   [arXiv:1203.3642 [hep-lat]].

\bibitem{Iritani:2018sra} 
  T.~Iritani {\it et al.},
  Phys.\ Lett.\ B {\bf 792}, 284 (2019)
    [arXiv:1810.03416 [hep-lat]].

\bibitem{Gongyo:2017fjb} 
  S.~Gongyo {\it et al.},
  Phys.\ Rev.\ Lett.\  {\bf 120},  212001 (2018)
  [arXiv:1709.00654 [hep-lat]].

\bibitem{Aoki:2011gt}
  S.~Aoki {\it et al.} [HAL QCD Collaboration],
  Proc.\ Japan Acad.\ B {\bf 87}, 509 (2011)
  [arXiv:1106.2281 [hep-lat]].


\bibitem{Aoki:2012bb}
  S.~Aoki, B.~Charron, T.~Doi, T.~Hatsuda, T.~Inoue and N.~Ishii,
  Phys.\ Rev.\ D {\bf 87}, 034512 (2013)
  [arXiv:1212.4896 [hep-lat]].


\bibitem{Garcilazo}
H.~Garcilazo, Phys.\ Rev.\ C {\bf 93}, 024001 (2016);
H. Garcilazo and A. Valcarce, Phys. Rev. C{\bf 93}, 034001 (2016).

\bibitem{Kamada:2001tv}
  H.~Kamada {\it et al.},
  Phys.\ Rev.\ C {\bf 64}, 044001 (2001)
  [nucl-th/0104057].


\bibitem{Nemura05} 
H. Nemura, S. Shinmura, Y. Akaishi, and Khin Swe Myint,
Phys. Rev. Lett. {\bf 94}, 202502 (2005).

\bibitem{Ahn01} 
J. K. Ahn, {\it et al.},  Phys. Rev. Lett. {\bf 87}, 132504 (2001).

\bibitem{Cottenssi}

L.~Contessi, M.~Sch$\ddot{\rm a}$fer, N.~Barnea, A.~Gal and J.~Mare$\check{\rm s}$,
Phys. Lett. B {\bf 797},  134893 (2019).


\end{thebibliography}
\end{document}